\documentclass[10pt,journal,compsoc]{IEEEtran}
\usepackage{cite}
\usepackage{amsmath,amssymb,amsfonts}
\usepackage{graphicx}
\usepackage{textcomp}
\usepackage{xcolor}
\usepackage{threeparttable}
\usepackage{tabularx}
\usepackage{booktabs}
\usepackage{multirow}
\usepackage{caption}
\usepackage{colortbl}
\usepackage{subfigure}
\usepackage{pifont}
\usepackage[square,numbers,sort&compress]{natbib}
\usepackage{amsthm}
\usepackage{fancyhdr}
\usepackage[normalem]{ulem}
\usepackage[hyphens]{url}
\usepackage{color}
\usepackage{soul}

\DeclareMathAlphabet\mathbfcal{OMS}{cmsy}{b}{n}

\newcommand{\mat}[1]{\mathbf{#1}}

\usepackage[linesnumbered,boxed,ruled,vlined]{algorithm2e}

\def\BibTeX{{\rm B\kern-.05em{\sc i\kern-.025em b}\kern-.08em
    T\kern-.1667em\lower.7ex\hbox{E}\kern-.125emX}}

\begin{document}


\title{Triangle Counting Accelerations: From Algorithm to In-Memory Computing Architecture}

\author{Xueyan~Wang,~\IEEEmembership{Member,~IEEE,}
        Jianlei Yang, Yinglin Zhao, Xiaotao Jia, Rong Yin, Xuhang Chen, \\Gang Qu,~\IEEEmembership{Fellow,~IEEE,}
        and~Weisheng~Zhao,~\IEEEmembership{Fellow,~IEEE}
\IEEEcompsocitemizethanks{
\IEEEcompsocthanksitem Corresponding authors are Jianlei Yang and Weisheng Zhao. E-mail: jianlei@buaa.edu.cn, weisheng.zhao@buaa.edu.cn
\IEEEcompsocthanksitem X. Wang, Y. Zhao, X.Jia, X. Chen and W. Zhao are with MIIT Key Laboratory of Spintronics, School of Integrated Circuit Science and Engineering, Beihang University, Beijing 100191, China.
\IEEEcompsocthanksitem J. Yang is with the School of Computer Science and Engineering, BDBC, State Key Laboratory of Software Development Environment (NLSDE), Beihang University, Beijing 100191, China.
\IEEEcompsocthanksitem R. Yin is with the Institute of Information Engineering, Chinese Academy of Sciences, Beijing, China.
\IEEEcompsocthanksitem G. Qu is with the Department of Electrical and Computer Engineering and Institute for Systems Research, University of Maryland, College Park, MD 20742 USA.
}
}


\IEEEtitleabstractindextext{
\begin{abstract}

Triangles are the basic substructure of networks and triangle counting (TC) has been a fundamental graph computing problem in numerous fields such as social network analysis.
Nevertheless, like other graph computing problems, due to the high memory-computation ratio and random memory access pattern, TC involves a large amount of data transfers thus suffers from the bandwidth bottleneck in the traditional Von-Neumann architecture.
To overcome this challenge, in this paper, we propose to accelerate TC with the emerging processing-in-memory (PIM) architecture through an algorithm-architecture co-optimization manner.
To enable the efficient in-memory implementations, we come up to reformulate TC with bitwise logic operations (such as \texttt{AND}), and develop customized graph compression and mapping techniques for efficient data flow management.
With the emerging computational Spin-Transfer Torque Magnetic RAM (STT-MRAM) array, which is one of the most promising PIM enabling techniques, 
the device-to-architecture co-simulation results demonstrate that the proposed TC in-memory  accelerator  outperforms  the  state-of-the-art  GPU and FPGA accelerations by $12.2\times$ and $31.8\times$, respectively, and achieves a $34\times$ energy efficiency improvement over the FPGA accelerator.


\end{abstract}

}

\maketitle

\IEEEraisesectionheading{\section{Introduction}\label{sec:introduction}}

\IEEEPARstart{T}{riangle} counting (TC) counts the number of triangles in a given graph and it is an basic problem in graph computing.
TC problem is not hard but it is memory bandwidth intensive thus time-consuming. As a result, researchers from both academia and industry have proposed many TC acceleration methods ranging from sequential to parallel, single-machine to distributed, and exact to approximate.
From the computing hardware perspective, these acceleration strategies are generally executed on CPU, GPU or FPGA, and are based on Von-Neumann architecture \cite{tcReview,XiongTCCPGPU,XiongTCFPGA}.
However, due to the fact that most graph processing algorithms have low computation-memory ratio and high random data access patterns, there are frequent data transfers between the computational unit and memory components which consumes a large amount of time and energy, the existing acceleration approaches can only alleviate by parallelism while cannot resolve the bottleneck of data transfers. 

Through performing computation where the data resides, in-memory computing paradigm can save most of the off-chip data communication energy and latency by exploiting the large internal memory inherent bandwidth and inherent parallelism \cite{MutluDRAM,DBLP:conf/dac/LiXZZLX16}. As a result, in-memory computing has appeared as a viable way to carry out the computationally-expensive and memory-intensive tasks \cite{LiBingOverview,FanAligns}. 
This becomes even more promising when being integrated with the emerging non-volatile Spin-Transfer Torque Magnetic RAM (STT-MRAM) memory technologies. This integration offers fast write speed, low write energy, and high write endurance among many other benefits \cite{wang2018current,DBLP:journals/tvlsi/JainRRR18}.

However, compared to the traditional Von-Neumann computing architecture, in which the CPU has very powerful and complex computing capabilities and control capabilities, the relatively dispersed in-memory processing cores in the spin-based in-memory computing architecture are more suitable for processing tasks that has relatively simple types of calculations and simple control logic.
Due to such data transmission mode and computing characteristics of the in-memory computing architecture, traditional graph algorithms are often not well applied to in-memory computing.
In the literature, there have been some explorations on in-memory graph algorithm accelerations \cite{ChenHPCA,FanGraphs,WangYuASPDAC,QianMicro}.
As analyzed above, existing TC algorithms cannot be efficiently implemented in memory.
For example, the intersection-based ones cannot be directly implemented in memory, and the matrix multiplication-based ones involve complex arithmetic computations which require non-trivial design overheads while implemented in-memory. 
In addition, for large sparse graphs, efficient graph data compression and data mapping mechanisms are all critical for PIM accelerations. The existing data compression methods for sparse graph, such as compressed sparse column (CSC), compressed sparse row (CSR), and coordinate list (COO) \cite{ChenHPCA}, cannot be directly applied to in-memory computation either. 

In this paper, we propose and design the first {triangle counting in-memory accelerator}, called TCIM, that overcomes the above barriers through an algorithm-architecture co-optimization approach.
We find that the number of triangles in a given graph can be computed using only \texttt{AND} and \texttt{BitCount} operations. 
Once the problem has been framed in this form, it can be efficiently implemented in an in-memory manner.
The contributions of this paper can be summarized as follows.


\begin{itemize}

\item A hardware-friendly triangle counting method is proposed using bitwise logic operations.
Such reformulation of triangle counting is amenable to in-memory implementations.
\item We propose customized data slicing for efficient graph data compression, and graph data flow management strategies for mapping onto in-memory computation architectures.
\item To support in-memory TC accelerations, a sparsity-aware processing-in-memory architecture is proposed utilizing state-of-the-art STT-MRAM technology. We also develop a device-to-architecture co-simulation framework for validating the proposed strategies.

\end{itemize}


The rest of the paper is organized as follows.
Section~\ref{sec:preliminary} provides some preliminary knowledge of triangle counting and in-memory computing.
Section~\ref{sec:tc} introduces the proposed TC method with bitwise operations, and Section~\ref{sec:pimArch} elaborates sparsity-aware data management strategies.
{Section~\ref{sec:overallArch} introduces the overall PIM architecture.}
Section~\ref{sec:exper} demonstrates the experimental results and Section~\ref{sec:conclusion} concludes the paper.

\section{Preliminary}\label{sec:preliminary}

\subsection{Triangle Counting}

Triangle counting problem seeks to determine the number of triangles in a given graph. It is essential for analyzing networks and generally considered as the first fundamental step in calculating metrics such as clustering coefficient and transitivity ratio, as well as other tasks such as community discovery, link prediction, and Spam filtering \cite{tcReview}. 
For example, the commonly used social analysis algorithm, community discovery, gives the number of triangles in a social network to analyze which circles are more stable and have closer relationships. For a person's social circle, the more triangles there are, the stronger and closer his social relationship is. 
{For network science in biology and neuroscience, it is also found useful to demonstrate the self-optimization phenomenon in brain's neuronal networks \cite{yin2020network} and to control biological network \cite{yang2020controlling}.}
The sequential algorithms for TC can be classified into two groups. 

In the {matrix multiplication based algorithms}, a triangle is a closed path of length three, namely a path of three vertices begins and ends at the same vertex. If $\mat{A}$ is the adjacency matrix of graph $G$, $\mat{A}^3[i][i]$ represents the number of paths of length three beginning and ending with vertex $i$. Given that a triangle has three vertices and will be counted for each vertex, and the graph is undirected (that is, a triangle $i-p-q-i$ will be also counted as $i-q-p-i$), the number of triangles in $G$ can be obtained as $trace(\mat{A}^3)/6$, where $trace$ is the sum of elements on the main diagonal of a matrix. 

In the {set intersection based algorithms}, it iterates over each edge and finds common elements from adjacency lists of head and tail nodes.
A lot of CPU, GPU and FPGA based optimization techniques have been proposed \cite{tcReview,XiongTCCPGPU,XiongTCFPGA}. These works show promising results of accelerating TC, however, these strategies all suffer from the performance and energy bottlenecks brought by the significant amount of data transfers in TC. 

\subsection{In-Memory Computing with STT-MRAM}

In-Memory Computing efforts can be classified into two categories according to whether they target at application-specific computations \cite{DBLP:conf/isca/AhnYMC15,liu2015reno,ramasubramanian2014spindle} or general-purpose computations \cite{DBLP:journals/tvlsi/JainRRR18,DBLP:journals/cal/ChowdhuryHKZLLS18,DBLP:conf/dac/LiXZZLX16,kang2017memory,parveen2018hielm,chang2019dasm,zhao2019stt}.
{ReRAM has been extensively explored and used to implement matrix-vector multiplication for neural network accelerations, with the multi-bit storage property. 
Comparatively, STT-MRAM has higher write endurance, faster write speed, lower write energy, while it only has limited resistance difference between the distinct resistance states of MTJ \cite{wang2018current}. In particular, prototype STT-MRAM chip demonstrations and commercial MRAM products have been available by companies such as Everspin and TSMC.
As a result, STT-MRAM is widely used to implement bit-wise boolean operations for general-purpose in-memory computing paradigm \cite{DBLP:journals/tvlsi/JainRRR18,guo2021spintronics}.
In this paper, we focus on such general-purpose {PIM}, which can be widely used in various categories of applications.}

STT-MRAM stores data with magnetic-resistances instead of conventional charge based store and access. 
Due to this current sensing mechanism in STT-MRAM and the fact that current can be accumulated, STT-MRAM is able to realize logic functions conveniently.
This enables MRAM to provide inherent computing capabilities for bitwise logic with the core bit-cell and array structure of STT-MRAM remain unchanged, and only needs minor changes to peripheral circuitry (such as sensing circuitry to generate required sensing current)  \cite{DBLP:journals/tvlsi/JainRRR18}\cite{yang2018exploiting}.

\begin{figure}[t]
\centering
\includegraphics[width = 1.0\linewidth]{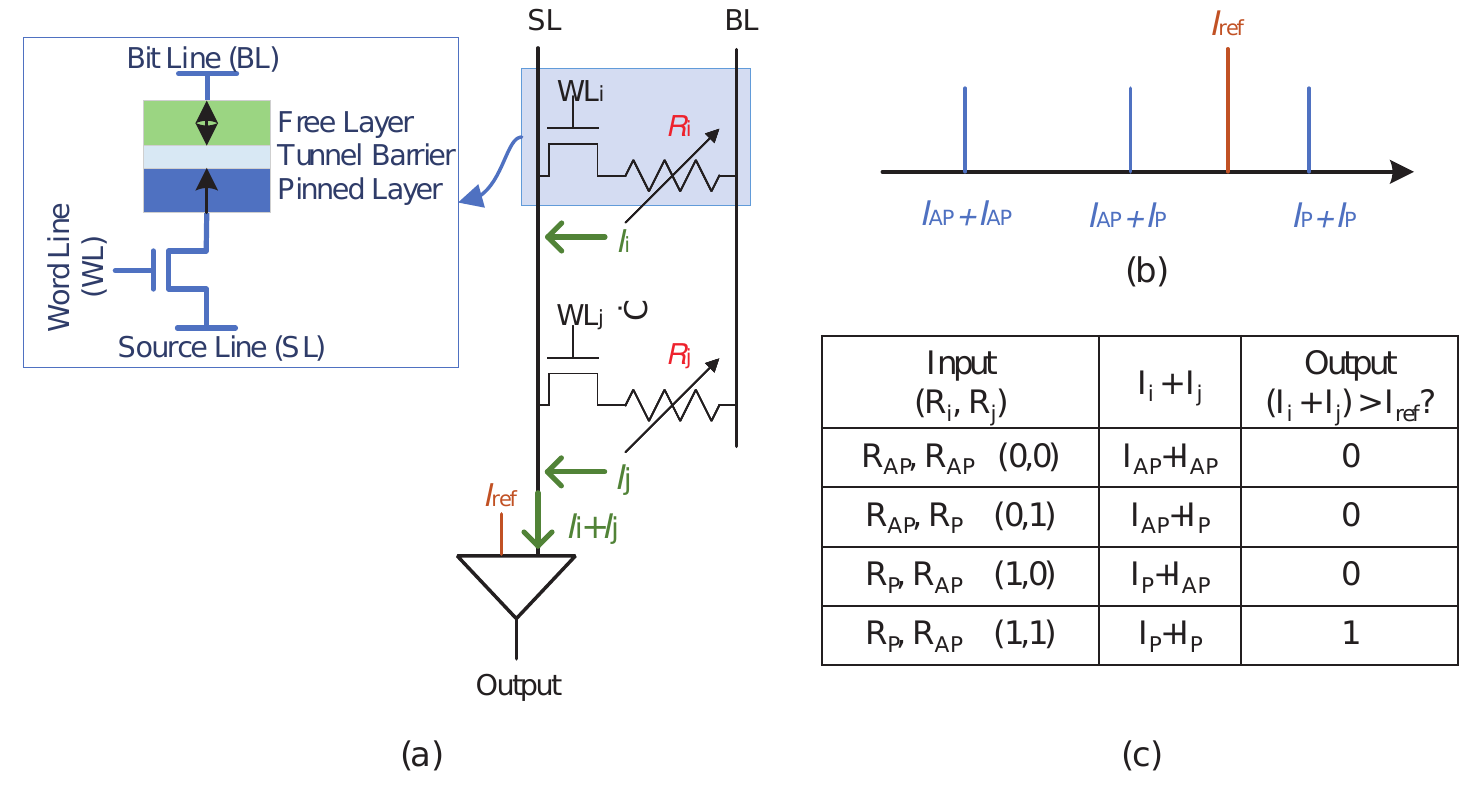}
\caption{An overview of performing Boolean \texttt{AND} operation with STT-MRAM. (a) Typical STT-MRAM bit-cells and computing paradigm. (b) The reference current. (c) Truth table.}
\label{fig:cim1}
\end{figure}

As Fig.~\ref{fig:cim1}(a) shows, a typical STT-MRAM bit-cell consists of an access transistor and a Magnetic Tunnel Junction (MTJ), which is controlled by bit-line (BL), word-line (WL) and source-line (SL).
The relative magnetic orientations of pinned ferromagnetic layer (PL) and free ferromagnetic layer (FL) can be stable in parallel (\texttt{P} state) or anti-parallel (\texttt{AP} state), corresponding to a low resistance ($R_{\rm P}$) or a high resistance ($R_{\rm AP}$), respectively. 
The \texttt{READ} operation is done by enabling WL signal, applying a voltage $V_{\rm read}$ across BL and SL, and sensing the current ($I_{\rm P}$ or $I_{AP}$) that flows through the MTJ. By comparing the sense current with a reference current ($I_{\rm ref}$,), the data stored in a MTJ cell (logic `0' or logic `1') could be readout.
The \texttt{WRITE} operation can be performed by enabling WL, then applying an appropriate voltage ($V_{\rm write}$) across BL and SL to pass a current that is greater than the critical MTJ switching current. 
To perform bitwise logic operation, by simultaneously enabling $WL_{\rm i}$ and $WL_{\rm j}$, then applying $V_{\rm read}$ across BL and SL, the current that feeds into the sense amplifier (SA) is a summation of $I_{\rm i}+I_{\rm j}$.
With different reference sensing current, various logic functions of the enabled word line can be implemented. For example, as shown in Fig.~\ref{fig:cim1}(b), when $I_{\rm ref} \in (I_{\rm AP}+I_{\rm P},I_{\rm P}+I_{\rm P})$, the truth table is demonstrated in Fig.~\ref{fig:cim1}(c), corresponds to \texttt{AND} logic.

\begin{figure}[t]
\centering
\includegraphics[width = 0.85\linewidth]{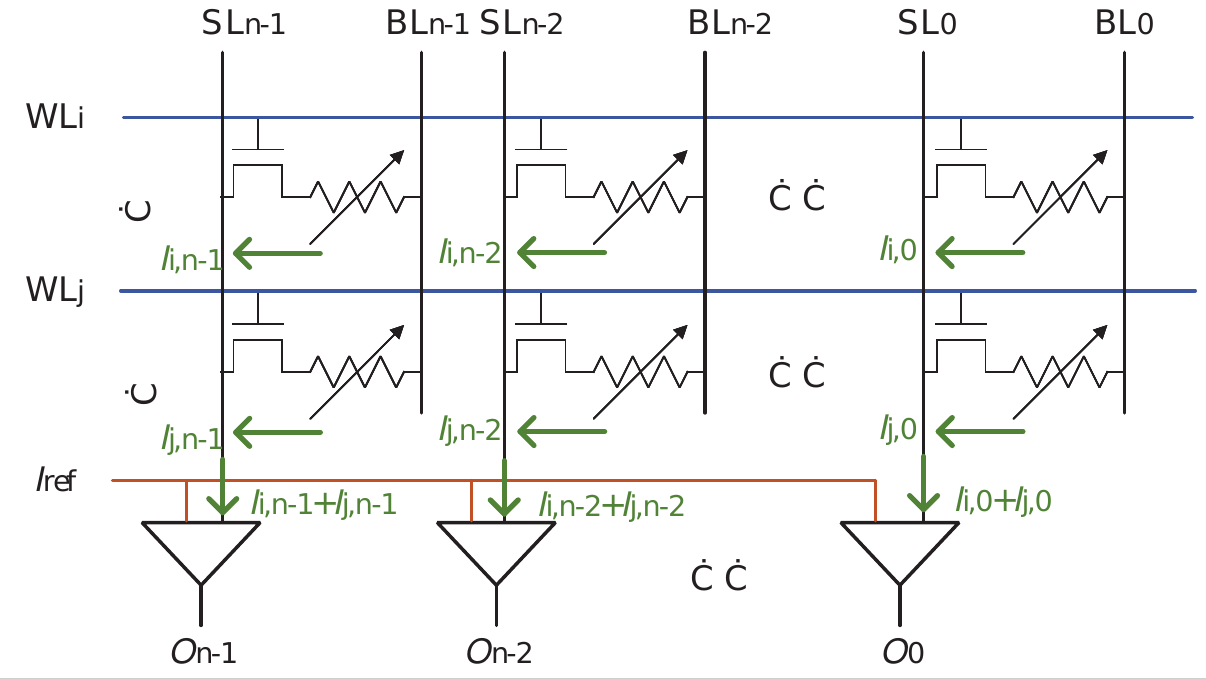}
\caption{Computational STT-MRAM array.}
\label{fig:cim2}
\end{figure}

Fig.~\ref{fig:cim2} demonstrates the STT-MRAM arrays that support in-memory logic computations. By simultaneously enabling word-line $\rm WL_i$ and $\rm WL_j$, then applying $V_{\rm read}$ across $\rm BL_k$ and $\rm SL_k$ ($\rm k \in [0,n-1]$), the current that feeds into the $\rm k$-th SA is a summation of the currents flowing through $\rm MTJ_{i,k}$ and $\rm MTJ_{j,k}$, namely $I_{\rm i,k}+I_{\rm j,k}$.
With different reference sensing current, the sense amplifier will have different outputs under given input patterns (corresponds to the high/low resistive state of the MTJs), then different logic functions of the enabled word line can be implemented.

\section{Reformulation of Triangle Counting}\label{sec:tc}

\begin{figure*}[t]
\centering
\includegraphics[width = 0.9\linewidth]{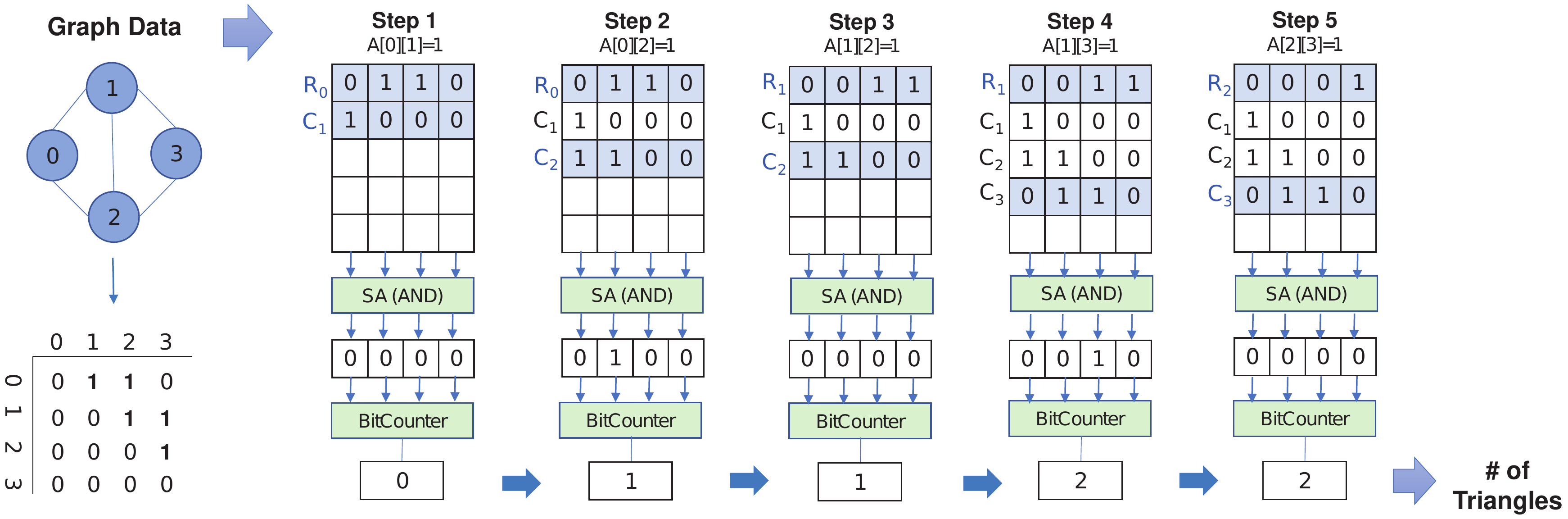}
\caption{Demonstrations of triangle counting with \texttt{AND} and \texttt{BitCount} bitwise operations.}
\label{fig:TCProcedure}
\end{figure*}

In this section, we seek to perform TC with massive bitwise operations, which is the enabling technology for in-memory TC accelerator.

\subsection{Triangle Counting with Bitwise Operations}\label{subsec:reformulatedTC}

Let $\mat{A}$ be the adjacency matrix representation of an undirected graph $G(V,E)$, where $\mat{A}[i][j]\in \{0,1\}$ indicates whether there is an edge between vertices $i$ and $j$.
If we compute $\mat{A}^2=\mat{A}*\mat{A}$, then the value of $\mat{A}^2[i][j]$ represents the number of distinct paths of length two between vertices $i$ and $j$.

In the case that there is an edge between vertex $i$ and vertex $j$ ($\mat{A}[i][j] \neq 0$), at the same time $i$ can also reach $j$ through a path of length two ($\mat{A}^2[i][j] \neq 0$), where the intermediate vertex is $k$, then vertices $i$, $j$, and $k$ form a triangle.
As a result, the number of triangles in $G$ is equal to the number of non-zero elements ($nnz$) in $\mat{A} \circ \mat{A}^2$ (the symbol `$\circ$' defines element-wise product), namely
\begin{equation}\label{equ:eq1}
  TC(G)=nnz(\mat{A} \circ \mat{A}^2).
\end{equation}
Since $\mat{A}[i][j]$ is either zero or one, we have 
\begin{equation}\label{equ:eq2}
  (\mat{A}\circ \mat{A}^2)[i][j]=
  \begin{cases}
    0, & \text{if}\ \mat{A}[i][j]=0;\\
    \mat{A}^2[i][j], & \text{if}\ \mat{A}[i][j]=1.
  \end{cases}
\end{equation}
According to Equation~(\ref{equ:eq2}),
\begin{equation}\label{equ:eq3}
\begin{split}
  nnz(\mat{A} \circ \mat{A}^2)&=\sum\sum\nolimits_{\mat{A}[i][j]=1}\mat{A}^2[i][j].
\end{split}
\end{equation}
Because the element in $\mat{A}$ is either zero or one, the bitwise Boolean \texttt{AND} result is equal to that of the mathematical multiplication, thus

\begin{equation}\label{equ:eq4}
\begin{split}
  \mat{A}^2[i][j]& =\sum_{k=0}^{n} \mat{A}[i][k]*\mat{A}[k][j]=\sum_{k=0}^{n} {AND}(\mat{A}[i][k],\mat{A}[k][j])\\
 & ={BitCount}({AND}(\mat{A}[i][*],\mat{A}[*][j]^T)),
\end{split}
\end{equation}
in which \texttt{BitCount} returns the number of `1's in a vector consisting of `0' and `1', for example, $BitCount(0110)=2$. 

Combining equations ~(\ref{equ:eq1}), (\ref{equ:eq3}) and (\ref{equ:eq4}), we have 
\begin{equation}
\begin{split}
  TC(G)&={BitCount}({AND}(\mat{A}[i][*],\mat{A}[*][j]^T)),\\
  &\quad s.t.~ \mat{A}[i][j]=1.
  \end{split}
\end{equation}

Therefore, TC can be completed by only \texttt{AND} and \texttt{BitCount} operations (massive for large graphs). 
For each non-zero entry in the adjacency matrix, the corresponding row and column are loaded into STT-MRAM computational memory where each cell consists of one transistor and one MTJ. 
Consequently, the \texttt{AND} computations are carried out within the STT-MRAM memory, and the bit counter is incremented by the number of $1$s in the result of the \texttt{AND} computations. 
The bit counter will eventually store the total number of triangles in the graph. 

\subsection{An Illustrative Example}

With the reformulated triangle counting method in Section~\ref{subsec:reformulatedTC}, for each non-zero element $\mat{A}[i][j]=1$, the $i$-th row ($R_i=\mat{A}[i][*]$) and the $j$-th column ($C_j=\mat{A}[*][j]^T$) are executed \texttt{AND} operation, then the \texttt{AND} result is sent to a bit counter module for accumulation. 
Once all the non-zero elements are processed, the value in the accumulated \texttt{BitCount} is the number of triangles in the graph.
Fig.~\ref{fig:TCProcedure} demonstrates an illustrative example.
The graph has four vertices and five edges, and the adjacency matrix is given. 
The non-zero elements in the adjacency matrix $\mat{A}$ are $\mat{A}[0][1]$, $\mat{A}[0][2]$, $\mat{A}[1][2]$, $\mat{A}[1][3]$, and $\mat{A}[2][3]$.

\begin{enumerate}

\item For $\mat{A}[0][1]$, row $R_0$='0110' and column $C_1$=`1000' are executed with \texttt{AND} operation, then the \texttt{AND} result `0000' is sent to the bit counter and gets a result of zero;

\item For $\mat{A}[0][2]$, row $R_0$=`0110' and column $C_2$=`1100' are executed with \texttt{AND} operation and the result is `0100', then the \texttt{BitCount} result of `0100' is one;

\item For $\mat{A}[1][2]$, row $R_1$=`0011' and column $C_2$=`1100' are executed with \texttt{AND} operation, then the \texttt{AND} result `0000' is sent to the bit counter, thus the result remains to be one;

\item For $\mat{A}[1][3]$, row $R_1$=`0011' and column $C_3$=`0110' are executed with \texttt{AND} operation, then the \texttt{AND} result `0010' is executed with \texttt{BitCount}, the bit counter is incremented by one, thus gets a result of two;

\item For $\mat{A}[2][3]$, row $R_2$=`0001' and column $C_3$=`0110' are executed with \texttt{AND} operation, then the \texttt{AND} result `0000' is sent to the bit counter, thus the result remains to be two.

\end{enumerate}

After the process of the last non-zero element $\mat{A}[2][3]$ is finished, the accumulated \texttt{BitCount} result is two, as a result, the graph has two triangles (corresponds to triangles "$0-1-2-0$" and "$1-2-3-1$" in the graph).

\subsection{Discussions on the Reformulated TC}

We show that the number of triangles in a given graph can be computed using only \texttt{AND} operations and bit counters on the adjacency matrix of the graph. Once the problem has been framed in this form, the method proposed for triangle counting looks at every non-zero entry in the adjacency matrix and, for each such entry, the corresponding row and column are loaded into STT-MRAM memory where each cell consists of one transistor and one MTJ. \texttt{AND} computations are then carried out within memory and a bit counter is incremented by one if the result of the computations is 1. The bit counter will eventually store the total number of triangles in the graph.

The proposed TC method has the following characteristics:
First, it avoids the traditional time-consuming matrix multiplications. Through making the operation data be either zero or one, we can simply implement the original multiplication with Boolean \texttt{AND} logic.
Second, the proposed method does not need to store the intermediate results that are larger than one (such as the elements in $\mat{A}^2$), which enables high storage efficiency and in-memory computation regularity.
Third, it does not need complex control logic. It only needs to iterate the non-zero elements and conduct corresponding \texttt{AND} and \texttt{BitCount} operations.


Given the above three characteristics, and the fact that in-memory computation is suitable for data-intensive applications with relative simple computation and control logic,
the proposed reformulated TC method is amenable to highly efficient in-memory computing structure.

\section{Sparsity-aware Graph Data Management for In-Memory Accelerations}\label{sec:pimArch}

Given that the size of the computational memory array is limited, and that most graph are highly sparse, efficient data flow management is critical for TC accelerations in order to reduce the unnecessary memory and computation requirements.
In this part, we will discuss about the data flow management techniques, including the data reuse/replacement and data compression methods, to minimize the needed memory space and computations when being mapped onto the computational memory array.

\subsection{Graph Data Reuse and Replacement}

The proposed TC method in Section~\ref{sec:tc} iterates over each non-zero element $\mat{A}[i][j]$ in the adjacency matrix $\mat{A}$, and loads its corresponding row $R_i$ and column $C_j$ into computational memory for \texttt{AND} operation.
As a result, all the non-zero elements in row $R_i$ can reuse this row for computations, and similarly, the non-zero elements in column $C_j$ can reuse this column. 
We propose data reuse strategy based on this observation.

Without loss of generality, we assume that the non-zero elements are iterated by rows.
For each processed row, it needs to be first loaded into the computational memory, then the corresponding columns of the non-zero elements in this row are sequentially loaded for \texttt{AND} computation.
In this case, once the computations for all the non-zero elements in a row have been finished, this row will no longer be used in future computations, thus this row can be overwritten by the next to-be-processed row. 
On the contrary, the corresponding columns might be used again while processing the non-zero elements in other rows.
As a result, before loading a certain column into memory for computation, we will first check whether this column has been loaded in previous computations. If it has existed in the computational memory, then it can be reused and save a memory \texttt{WRITE} operation, and if not, the column will be loaded to a spare computational memory space.
Overlapping the rows and reusing the columns can effectively reduce unnecessary space utilization and memory \texttt{WRITE} operations.

Here remains two questions to be answered:

\begin{itemize}
    \item First, how to decide the row sequence of processing?
    \item Second, in case that the computational memory is full, by what data replacement policy to swap data?
\end{itemize}

On selecting the next to-be-processed row, in a greedy way, the local optimal strategy is to choose the next row that has maximum overlaps with the current row on the columns of 1's, and in the ideal case, all the columns should be data hit. 
However, in case that the size of the matrix is huge then the columns of 1's may not be able to fit in the computational memory. Also, finding the row that overlaps most with the current row will increase non-trivial computational effort.
Alternatively, one may do in the zig-zag way: the first row goes from left to right to load the columns with 1, the second row goes from right to left to reuse the columns that are already in. This zig-zag way will work well in case of dense graphs.
As for the highly sparse graphs, we will simply process each row sequentially in the order in which the graph data is stored.

Take the case in Fig.~\ref{fig:TCProcedure} as an example, in step $1$ and step $2$, the two non-zero elements $\mat{A}[0][1]$ and $\mat{A}[0][2]$ are processed respectively, and corresponding row $R_0$ and columns $C_1$ and $C_2$ are loaded to memory. 
Next, while processing $\mat{A}[1][2]$ and $\mat{A}[1][3]$, $R_1$ will overlap $R_0$ and reuse existing $C_2$ in step $3$, and load $C_3$ in step $4$. 
In step $5$, to process $\mat{A}[2][3]$, $R_1$ will be overlapped by $R_2$, and $C_3$ is reused. 

For the data replacement policy, when the computational memory is full and a new column needs to be loaded into the memory for computation, we need to select one candidate column to be swapped out. {We know that a good data replacement algorithm should have a low replacement frequency, as a result, data that will not be accessed in the future or will not be accessed for a long time in the future should be swapped out first.}

{According to our proposed TC method, we need to iterate the non-zero elements in the adjacency matrix by rows. On iterating one certain non-zero element, we need to load the corresponding column for \texttt{AND} and \texttt{BitCount} computations. At the same time, we will also record which columns have been loaded into the computational memory.
Therefore, we are able to know about the future computations and the storage status in the computational memory array. Given the above information, when the computational memory is full and a data replacement happens, we are able to locate the column in the computational memory with the longest time between the next visit and swap it out. }

{
Compared to the traditional data replacement strategies such as the LRU (Least Recently Used) policy, which predicts a good choice on choosing to-be-swapped candidate column according to the past computations, our proposed method is able to make the optimal decision with the knowledge of future executions.
This can cause the least data replacement frequency, and we name it as Priority data replacement policy.
}





\subsection{Graph Data Compression}\label{subsec:dataCompression}

To utilize the sparsity of the graph to reduce the memory requirement and unnecessary computation, we propose a data slicing strategy for graph data compression.

Assume $R_i$ is the $i$-th row, and $C_j$ is the $j$-th column of the adjacency matrix $\mat{A}$ of graph $G(V,E)$. 
Let the slice length be $|S|$ (namely each slice contains $|S|$ bits), then each row and column has $\lceil\frac{|V|}{|S|}\rceil$ slices.
Accordingly, the $k$-th slice in row $R_i$, which is represented as $R_i S_k$, can be formulated as 
\[R_i S_k = \{\mat{A}[i][k*|S|],\cdots,\mat{A}[i][(k+1)*|S|-1]\}.\]

We define slice $R_i S_k$ is \textbf{\textit{valid}} if and only if it has at least one non-zero element, namely \[\exists \mat{A}[i][t] \in R_i S_k,\mat{A}[i][t]=1,t\in [k*|S|,(k+1)*|S|-1].\]
Similar for the the $k$-th slice of column $C_j$:
\[C_j S_k = \{\mat{A}[k*|S|][j],\cdots,\mat{A}[(k+1)*|S|-1][j]\}.\]
Slice $C_j S_k$ is \textbf{\textit{valid}} if and only if \[\exists \mat{A}[t][j] \in C_j S_k,\mat{A}[t][j]=1,t\in [k*|S|,(k+1)*|S|-1].\]

Recall that for each non-zero element $\mat{A}[i][j]=1$ in the adjacency matrix, we need to compute the \texttt{AND} of its corresponding row $R_i$ and column $C_j$. 
With the proposed row and column slicing methods, we will perform the \texttt{AND} operation in the unit of slices, and we only need to process the valid slice pairs. Namely only when both of the row slice $R_i S_k$ and column slice $C_j S_k$ are valid, we will load the valid slice pair $(R_iS_k,C_jS_k)$ into the computational memory array for \texttt{AND} operation.

\begin{figure}[t]
\centering
\includegraphics[width = 1.0\linewidth]{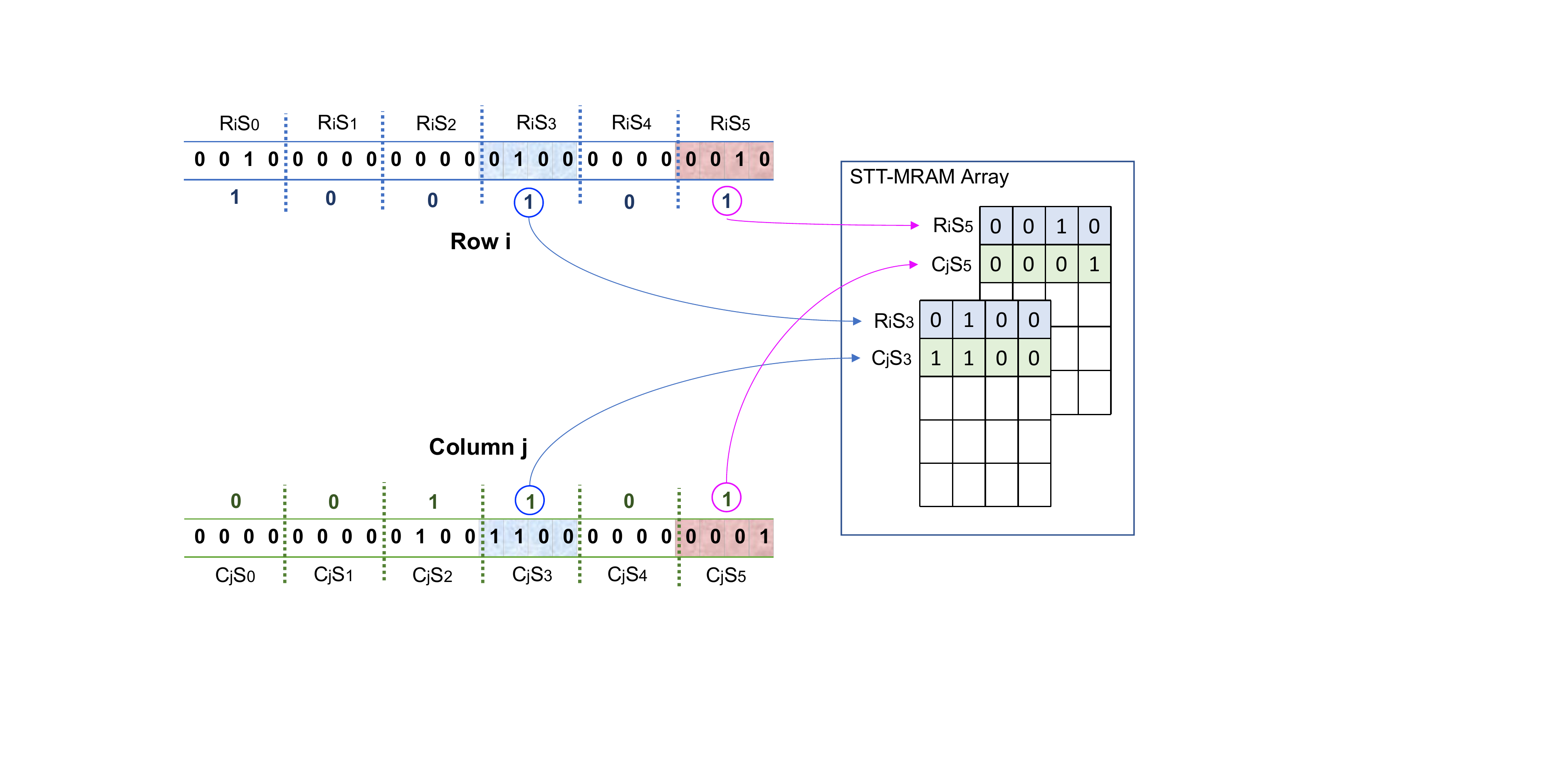}
\caption{Sparsity-aware data slicing and mapping.}
\label{fig:rowslicing}
\end{figure}

Fig.~\ref{fig:rowslicing} demonstrates an example, after row and column slicing, 
only the valid slice pairs $(R_iS_3,C_jS_3)$ and $(R_iS_5,C_jS_5)$ will be enabled for \texttt{AND} computation. 
This gives a glance of the fact that this filter process can reduce the needed computation significantly, especially in the large sparse graphs.

\textbf{Compression rate analysis.}
Assume that the graph has $|V|$ nodes, $|E|$ edges, the slice length is $|S|$, 
the sparsity of $G$ is defined as 
\[\alpha=1-\frac{|E|}{|V|^2}.\]

Therefore, $\alpha$ intuitively demonstrates the probability for an element in the adjacency matrix to be zero.
Accordingly, the probability for a slice with length of $|S|$ to be invalid (all elements in the slice should be zero) is 
$\alpha^{|S|}$.
Correspondingly, the probability for a slice to be valid (at least one element in the slice should be non-zero) is 
$1-\alpha^{|S|}$.
The number of valid slices $N_{VS}$ can be formulated as:
\[N_{VS}=(1-\alpha^{|S|}) \cdot \frac{|V|^2}{|S|}.\]

For data compression, we need to store the index of valid slices and the detailed data information of these slices. 
Assume that we need $|D|$ bits to store the index of slice ($|D| \geq log_2{\frac{|V|}{|S|}}$),
%
then the overall needed space (in Bytes) for compressed graph $G$ is 
\begin{figure*}[!ht]
\centering
\includegraphics[width = 0.9\linewidth]{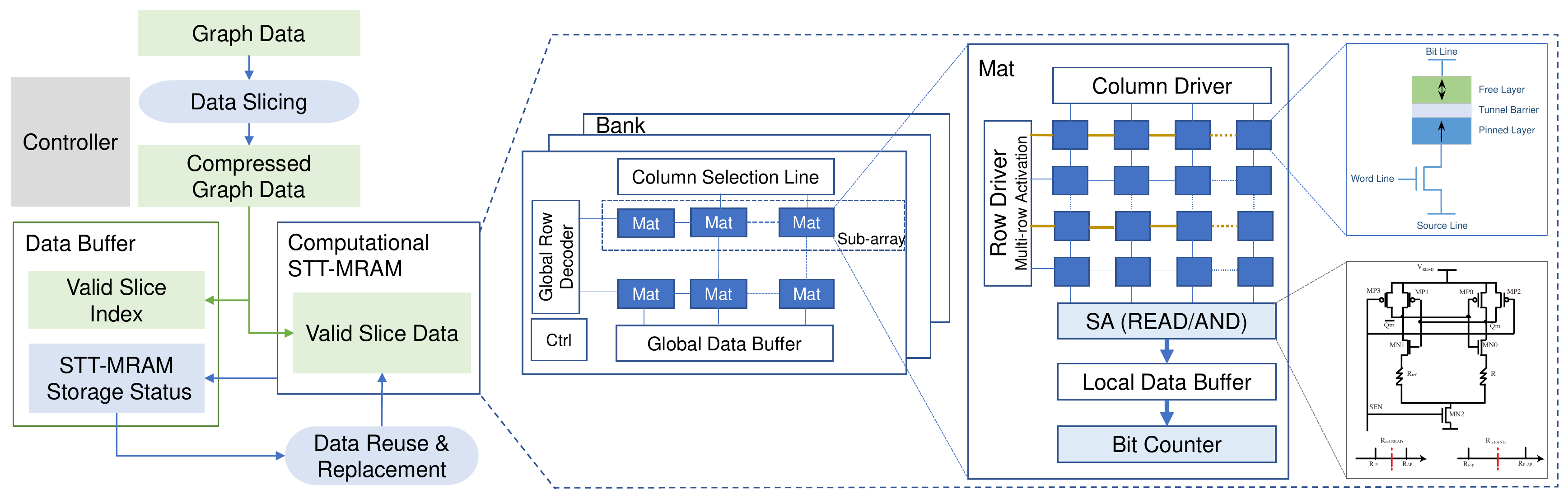}
\caption{Overall processing-in-memory architecture.}
\label{fig:overallArch}
\end{figure*} 

\begin{align*}
\text{Compressed Graph Size}  &= N_{VS} \times (\frac{|D|+|S|}{8}) \\
&=(1-\alpha^{|S|}) \cdot \frac{|V|^2}{|S|} \cdot (\frac{|D|+|S|}{8}).
\end{align*}

Without data slicing and compression, the needed storage space (in Bytes) is 
\[\text{Ordinary Graph Size} = \frac{|V|^2}{8}.\]

Consequently, the compression rate of the graph data can be expressed as:
\begin{align*}
\text{Compression Rate}~\mathit{CR} &= \frac {\text{Compressed Graph Size}}{\text{Ordinary Graph Size}} \\
&=(1+\frac{|D|}{|S|}) \cdot (1-\alpha^{|S|})
\end{align*}

\begin{figure}[h]
  \centering
    \subfigure[$\alpha\in (0,1)$, $|D|=32$]{\includegraphics[width=0.24\textwidth]{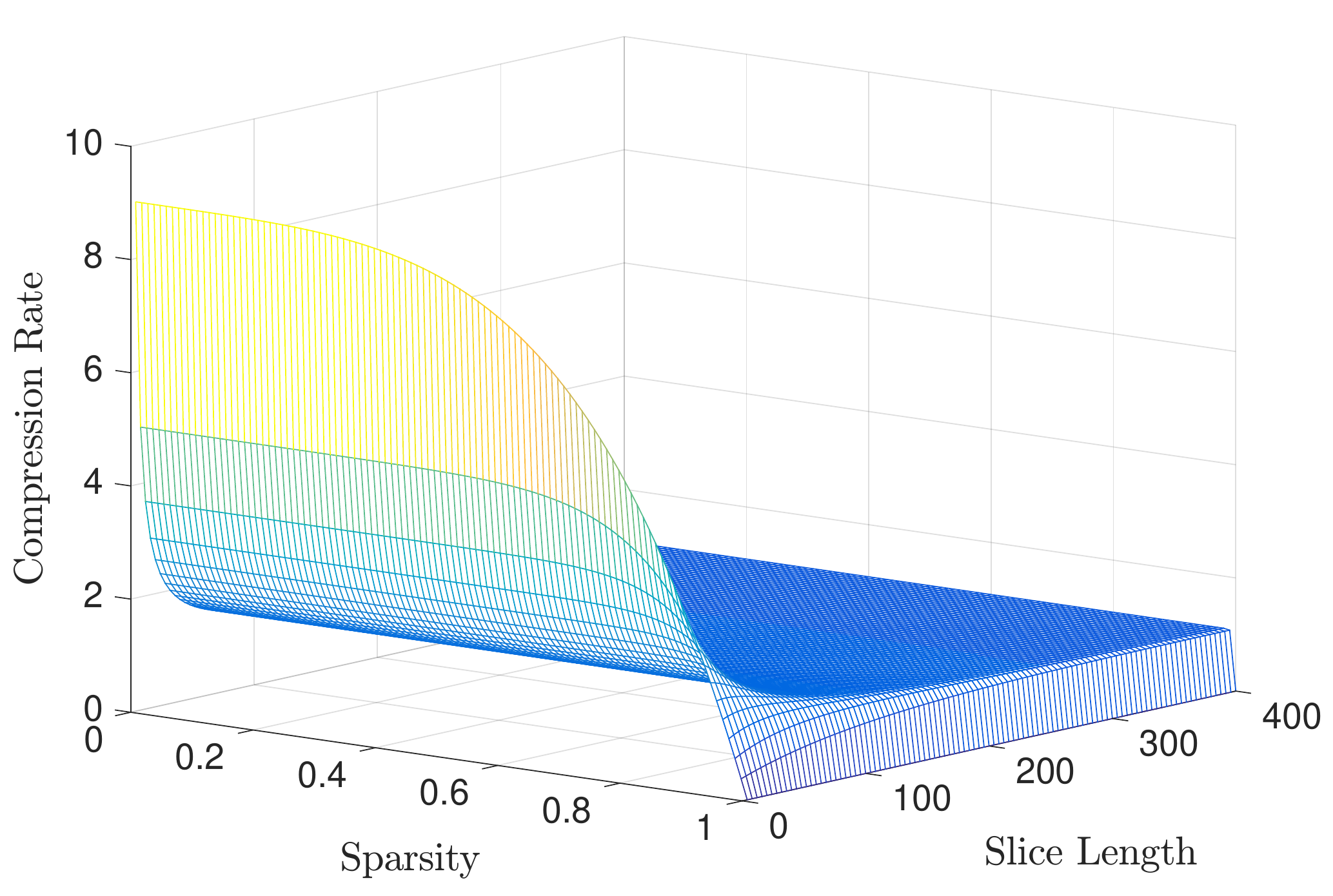}}
	\subfigure[$\alpha\in (0.9,1)$, $|D|=32$]{\includegraphics[width=0.24\textwidth]{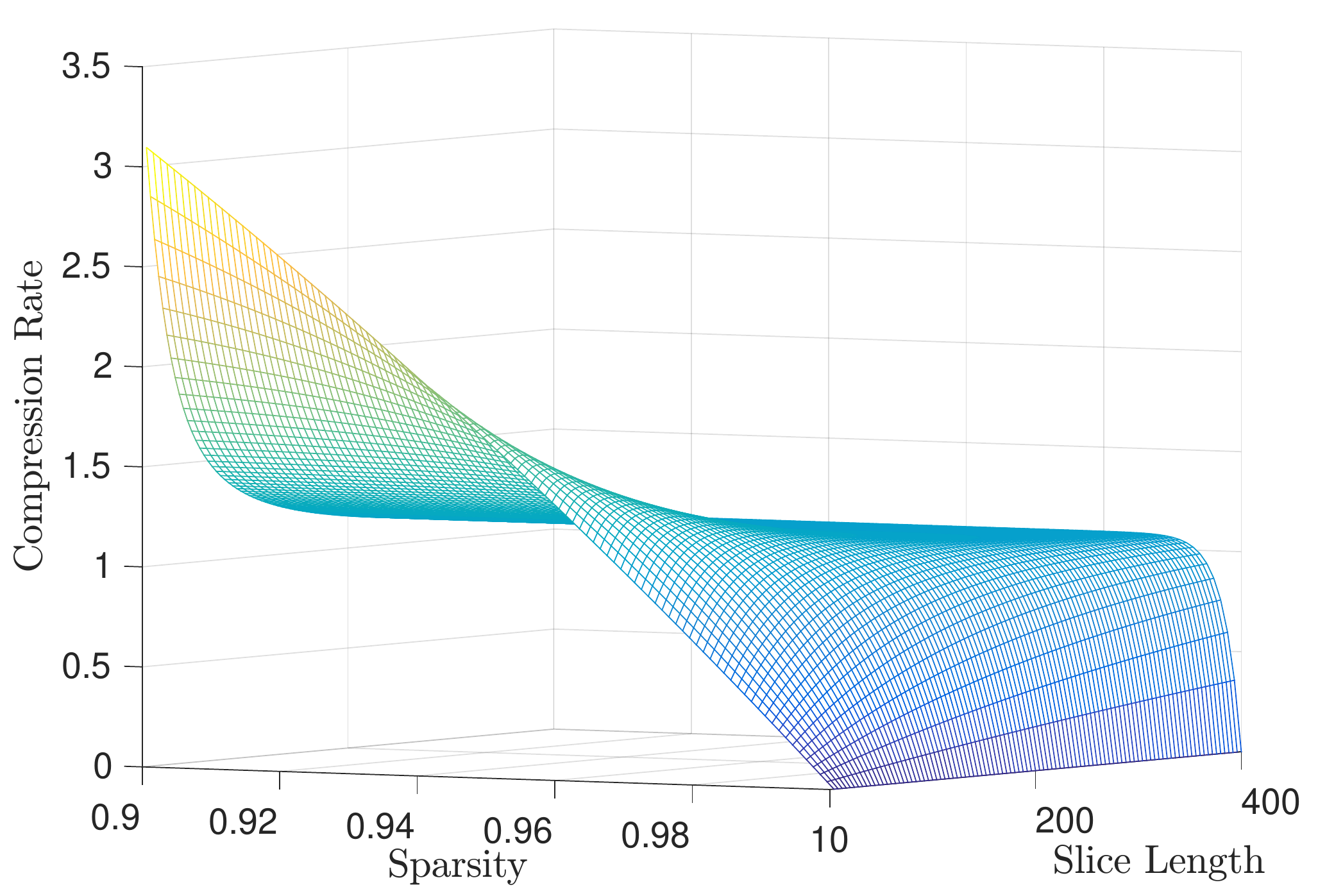}}
  \caption{Compression rate with different sparsity and slice/index length.}
	\label{fig:compressionRate}
\end{figure}

Therefore, the graph compression rate is determined by the sparsity of the graph, the slice length and the graph size. Fig.~\ref{fig:compressionRate}(a) demonstrates the compression rate with different graph sparsity and slice length when we use an integer ($|D|=32$) to store each valid slice index, and Fig.~\ref{fig:compressionRate}(b) zooms in the figure when the sparsity $\alpha \in (0.9,1)$.
We can see that the graph compression rate is dominated by the graph sparsity, when the sparsity is larger than 0.99, the compression method is expected to have a high compression efficiency.
Given that most graphs are highly sparse, the needed space to store the graph can be trivial and the experimental section will demonstrate some results.

\textbf{More importantly, the proposed format of compressed graph data is friendly for directly mapping onto the computational memory arrays to perform in-memory logic computation.} This is because the proposed compression method does not compress the valid slice data, thus does not need complex decompression process.

\section{Overall Architecture and Implementation}\label{sec:overallArch}

\subsection{Overall Architecture Design}

Fig.~\ref{fig:overallArch} demonstrates the overall architecture of the proposed TC accelerator.
First, the graph data will be sliced and compressed, and represented by the valid slice index and corresponding slice data. 
Consequently, according to the valid slice indexes in the data buffer, the corresponding valid slice pairs are loaded into computational STT-MRAM array for bitwise computation.
The storage status of STT-MRAM array (such as which slices have been loaded) is also recorded in the data buffer and utilized for data reuse and replacement. 

As for the computational memory array organization, each chip consists of multiple Banks and works as computational array. 
Each Bank is comprised of multiple computational sub-arrays, which are connected to a global row decoder and a shared global row buffer. Read circuit and write driver of the memory array are modified for processing bitwise logic functions. Specifically, the operation data are stored in different rows in memory arrays. The rows associated with operation data will be activated simultaneously for computing.
Sense amplifiers are enhanced with \texttt{AND} reference circuits to realize either \texttt{READ} or \texttt{AND} operations. 

{Note that in traditional Von-Neumann computing architecture, CPU is the central unit for control and computations, which can efficiently deal with complex computing and control task. In contrast, for in-memory processing, the decentralized processing cores can provide ultra-high parallelism, while they are more suitable for relatively single types of calculations with less control logic, such as the neural network computations. Data-intensive applications (such as the triangle counting graph algorithm demonstrated in this paper) which can be reformulated as simple logic computations are amenable to the proposed architecture.}

{Some nice work on graph computing accelerations has been proposed, such as GraphR \cite{ChenHPCA} and GraphSAR \cite{WangYuASPDAC}. They point out that large-scale graph calculation problems can be simulated in memristor array in the form of matrix vector operations. However, for graph computing, the implementation of vector matrix multiplication through analog operations faces the problem of accuracy and the additional overhead caused by digital-to-analog/analog-to-digital conversion. The proposed approach in this article reformulates the graph computing problem into basic Boolean logic functions, which can be implemented efficiently in-memory.}

\subsection{Pseudo-codes for In-Memory TC Acceleration}

\begin{algorithm}[t]
\caption{Triangle Counting with processing-in-memory Architecture.}
\label{alg:dataMapping}
\KwIn{Graph $G(V,E)$.}
\KwOut{The number of triangles in $G$.}
$TC\_G$ = 0\;
Represent $G$ with adjacent matrix $\mat{A}$\;
Iterate the non-zero elements in $\mat{A}$ by rows\;
\For {each non-zero element $\mat{A}[i][j]=1$}{
Partition $R_i$ into slices\;
Partition $C_j$ into slices\; 
\For {each valid slice pair ($R_iS_k$,$C_jS_k$)}{
$TC\_G$ += \textbf{COMPUTE} ($R_iS_k$,$C_jS_k$)\;
} 
}
\textbf{return} $TC\_G$ as the number of triangles in $G$.\\
----------------------------------------\\
\textbf{COMPUTE} ($RowSlice$, $ColumnSlice$)
{\\
Load $RowSlice$ into memory\;
\If {$ColumnSlice$ does not exist in the computational memory}{
\If {there is no enough space}
{
one slice with the longest time between the next visit to be swapped out;
}
Load $ColumnSlice$ into memory\;}
\textbf{return} $\texttt{BitCount}(\texttt{AND}(RowSlice,ColumnSlice))$.
}
\end{algorithm}

Algorithm~\ref{alg:dataMapping} demonstrates the pseudo-code for TC accelerations with the proposed architecture.
It iterates over each edge of the graph (corresponds to each non-zero element in the adjacency matrix) and partitions the corresponding rows and columns into slides, then loads the valid slice pairs onto computational memory for \texttt{AND} and \texttt{BitCount} computation. 
In case that there is no enough memory space, it will select one slice with the longest time between the next visit to be swapped out by the new slice.
Then repeat the above process until all the non-zero elements in the adjacency matrix is processed, and the accumulated \texttt{BitCount} result will be the number of triangles in the graph.

\section{Experimental Results}\label{sec:exper}

\subsection{Experimental Setup}

To validate the effectiveness of the proposed methods, comprehensive device-to-architecture evaluations along with two in-house simulators are developed. 

\begin{table}[b]
\setlength{\tabcolsep}{10pt}
\small
\caption{Key parameters for MTJ simulations.}
\label{tab:parameter}
\centering
\begin{tabular}{ll}
\specialrule{0.8pt}{0pt}{0pt}
 Parameter & Value \\ \hline
 MTJ Surface Length & $40$ $nm$ \\ 
 MTJ Surface Width & $40$ $nm$ \\ 
 Spin Hall Angle & $0.3$ \\ 
 Resistance-Area Product of MTJ & $10^{-12}$ $\Omega \cdot m^2$ \\ 
 Oxide Barrier Thickness & $0.82$ $nm$ \\ 
 TMR & $100\%$ \\ 
 Saturation Field & $10^6$ $A/m$ \\ 
 Gilbert Damping Constant & $0.03$ \\ 
 Perpendicular Magnetic Anisotropy & $4.5 \times 10^5$ $A/m$ \\
 Temperature & $300~K$ \\ 
\specialrule{0.8pt}{0pt}{0pt}
\end{tabular}
\end{table}

At the device level, we jointly use the Brinkman model and Landau-Lifshitz-Gilbert (LLG) equation to characterize MTJ \cite{yang2015radiation}. The key parameters for MTJ simulation are demonstrated in TABLE~\ref{tab:parameter}. 
For the circuit-level simulation, we design a Verilog-A model for 1T1R STT-MRAM device, and characterize the circuit with $45$nm FreePDK CMOS library. 
We design a bit counter module based on Verilog HDL to obtain the number of non-zero elements in a vector. Specifically, we split the vector and feed each $8$-bit sub-vector into an $8$-$256$ look-up-table to get its non-zero element number, then sum up the non-zero numbers in all sub-vectors. We synthesis the module with Synopsis Tool and conduct post-synthesis simulation based on $45$nm FreePDK.
{The modified sense amplifier part (to support logic computations) is also simulated in Cadence tool on $45$nm FreePDK.}
After getting the circuit-level simulation results, we integrate the parameters into the open-source NVSim simulator \cite{NVSim} and obtain the memory array performance {with wordwidth of 64 bits, 8-way cache configuration.}
In addition, we develop a simulator in Java for the processing-in-memory architecture, which simulates the proposed function mapping, data slicing and data mapping strategies. 
Finally, a behavioral-level simulator is developed in Java, taking architectural-level results and memory array performance to calculate the latency and energy that is spent on TC in-memory accelerator.

\begin{table*}[!h]
\setlength{\tabcolsep}{12pt}
\small
\caption{SNAP graph dataset.}
\label{tab:graphpara}
\centering
\begin{tabular}{lrrrl}
\specialrule{0.8pt}{0pt}{0pt}
 Graph & \# Vertices & \# Edges & \# Triangles & Description\\ \hline
 ego-facebook & 4039 & 88234 & 1612010 & Social circles from Facebook (anonymized)\\
 email-enron & 36692 & 183831 & 727044 & Email communication network from Enron\\
 com-Amazon & 334863 & 925872 & 667129 & Amazon product network\\ 
 com-DBLP & 317080 & 1049866 & 2224385 & DBLP collaboration network\\
 com-Youtube & 1134890 & 2987624 & 3056386 & Youtube online social network \\
 roadNet-PA & 1088092 & 1541898 & 67150 & Road network of Pennsylvania\\
 roadNet-TX & 1379917 & 1921660 & 82869 & Road network of Texas\\
 roadNet-CA & 1965206 & 2766607 & 120676 & Road network of California\\
 com-LiveJournal & 3997962 & 34681189 & 177820130 & LiveJournal online social network\\
\specialrule{0.8pt}{0pt}{0pt}
\end{tabular}
\end{table*}

To provide a solid comparison with other accelerators, we select from the real-life graphs from SNAP dataset \cite{snapnets} (see TABLE~\ref{tab:graphpara}), and run comparative baseline intersect-based algorithm on Inspur blade system with the Spark GraphX framework on Intel E5430 single-core CPU. For fair comparisons, our TC in-memory acceleration algorithm also runs on single-core CPU.

\subsection{Evaluations of Data Slicing and Compression}

\begin{figure}[t]
\centering
\includegraphics[width = 1.0\linewidth]{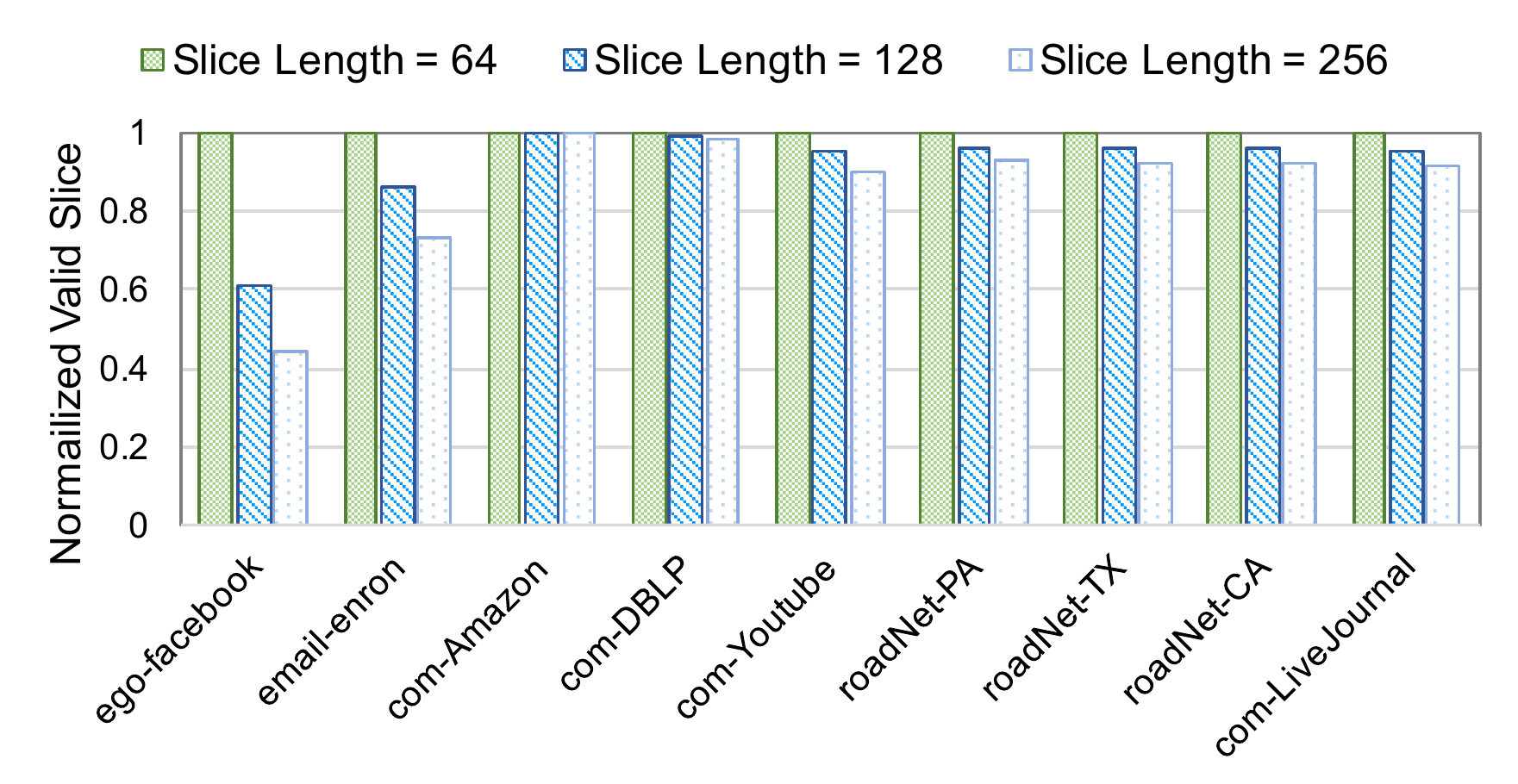}
\caption{The number of valid slices with the slice length being $64$, $128$ and $256$, respectively.}
\label{fig:dataslice}
\end{figure}

For the convenience and efficiency of computing, we can set the slice length to be the multiple of the computer word length. 
We assume the computer word length to be 64 bits in this paper.
Fig.~\ref{fig:dataslice} demonstrates the normalized valid slice number when the slice length are $64$, $128$, and $256$, respectively.
We can see that the number of valid slices only demonstrate a trivial reduction (on average less than $10\%$) when the slice length increase from 64 bits to $128$/$256$ bits (each slice has $2\times$/$4\times$ more bits).
Therefore, we set $|S|=64$ in the following experiments.

\begin{table}[t]
\setlength{\tabcolsep}{9pt}
\small
\caption{The sparsity of the graph dataset and the compression metrics by data slicing with slice length $|S|=64$ and index length $|D|=32$.}
\label{tab:dataSlice}
\begin{threeparttable}
\centering
\begin{tabular}{lccc}
\specialrule{0.8pt}{0pt}{0pt}
Graph & $\alpha$\tnote{$*$} & $CR$\tnote{$**$} & $VSR$\tnote{$\dagger$} \\ \hline
ego-facebook & 99.45914\% & 11.154\% & 7.017\% \\
email-enron & 99.98635\% & 0.584\% & 1.483\% \\
com-Amazon & 99.99917\% & 0.078\% & 0.014\% \\
com-DBLP & 99.99896\% & 0.080\% & 0.036\% \\
com-Youtube & 99.99977\% & 0.014\% & 0.013\% \\
roadNet-PA & 99.99987\% & 0.009\% & 0.013\% \\
roadNet-TX & 99.99990\% & 0.007\% & 0.010\% \\
roadNet-CA & 99.99993\% & 0.005\% & 0.007\% \\
com-LiveJournal & 99.99978\% & 0.013\% & 0.006\% \\
\specialrule{0.8pt}{0pt}{0pt}
\end{tabular}
\begin{tablenotes}
\item[$*$] Sparsity of the graph
\item[$**$] Compression rate
\item[$\dagger$] Valid slice pair ratio
\end{tablenotes}
\end{threeparttable}
\end{table}

TABLE~\ref{tab:dataSlice} demonstrates the sparsity of the each benchmark in the SNAP graph dataset and the corresponding compression rate when the slice length is $64$ and index length is $32$. 
As shown in the second and third columns of TABLE~\ref{tab:dataSlice}, the real-world graph are highly sparse, which leads to an extreme low compression rate, which validates the theoretical analysis in Section~\ref{subsec:dataCompression}.
As shown in the fourth column of TABLE~\ref{tab:dataSlice}, the valid slice pairs occupy a very small percentage among the whole slices, and this also leads to a high computation efficiency.
The average sparsity of the five largest graphs is $99.999\%$, with the average compression rate and average percentage of valid slices be $0.01\%$, This means the proposed data slicing and compression strategy could significantly reduce the needed memory space and computations by $99.99\%$.

\subsection{Evaluations of Data Reuse and Replacement}

 
\begin{figure}[b]
\centering
\includegraphics[width = 1.0\linewidth]{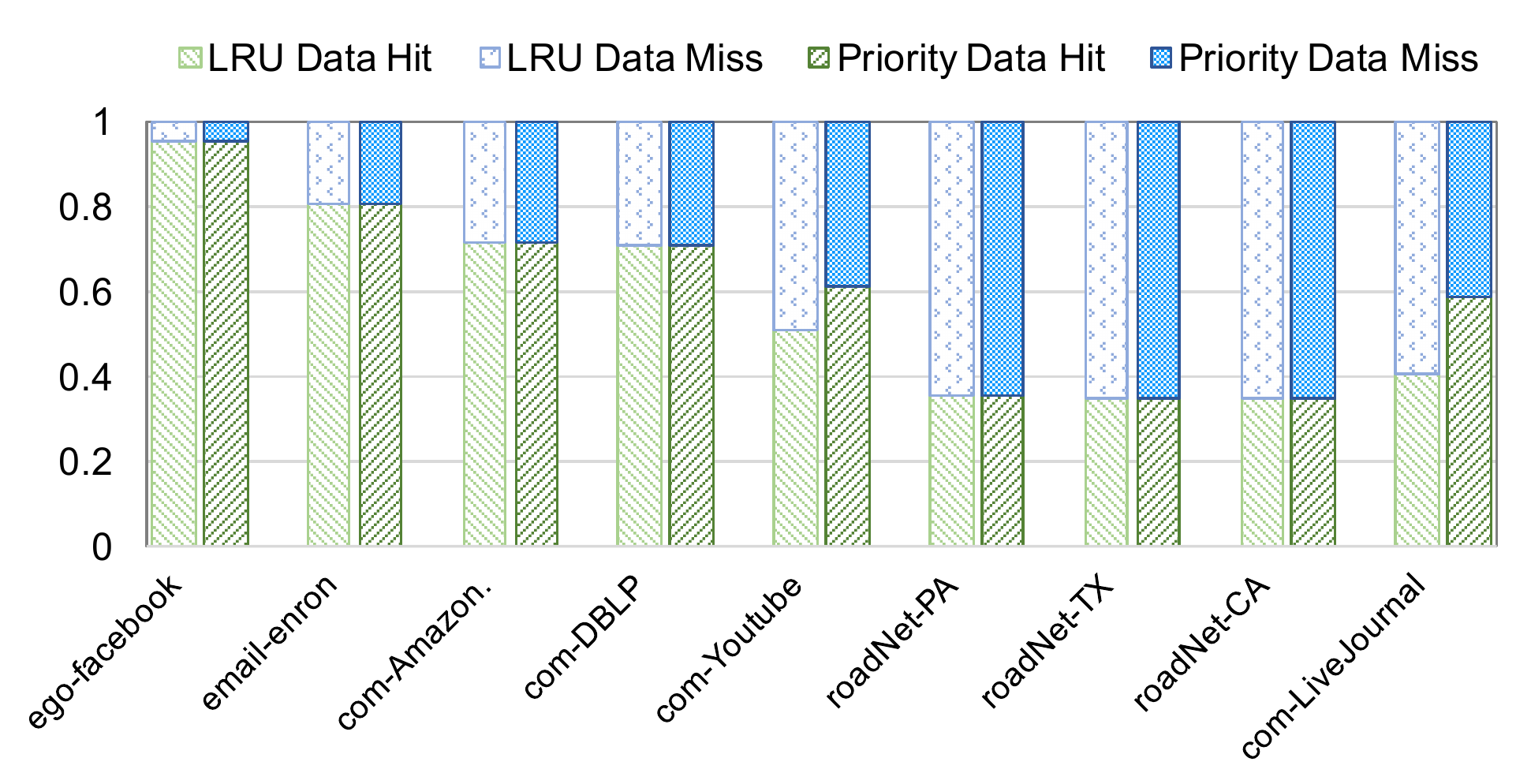}
\caption{Data hit and data miss ratios with LRU and Priority data replacement strategies.}
\label{fig:datareuse}
\end{figure} 

We know that the first time a data slice is loaded, it is always a miss, and a data hit implies that the slice data has already been loaded and a data reuse has happened. 
And when the required computational memory is larger than the STT-MRAM computational memory size, at the same time a data miss occurs, then data replacement will happen. 

With $8$ MB STT-MRAM computational memory array, in Fig.~\ref{fig:datareuse}, we have listed the ratios of data hit and data miss ratios under LRU and Priority data replacement policies. 
For the Priority data replacement policy, the data hit and data miss ratios are $60.5\%$ and $39.5\%$, respectively.
The data hit rate implies that the proposed data reuse strategy saves on average $60.5\%$ memory \texttt{WRITE} operations. 

\begin{figure}[t]
\centering
\includegraphics[width = 1.0\linewidth]{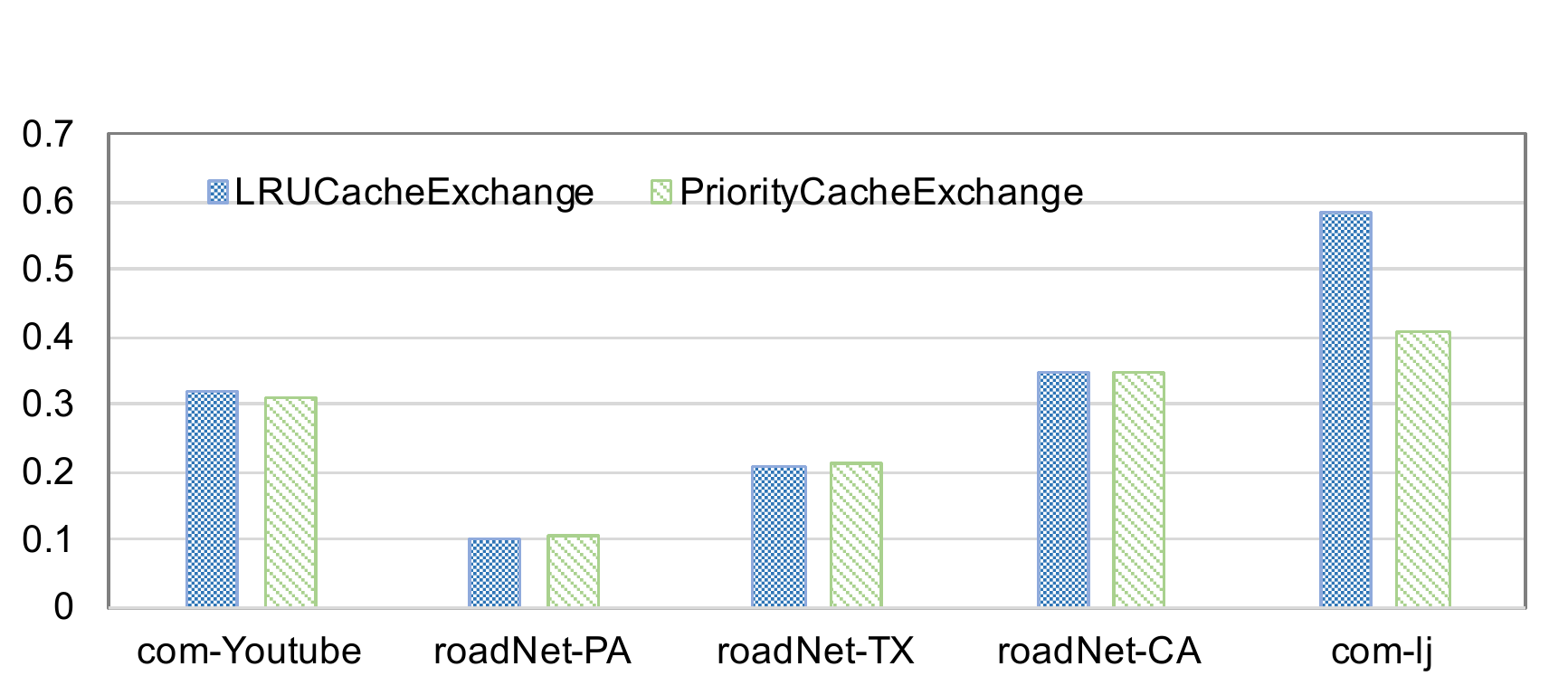}
\caption{Data replacement ratio with LRU and Priority data replacement strategies.}
\label{fig:dataexchange}
\end{figure}

The five largest graphs, including {\it com-Youtube}, {\it roadNet-PA}, {\it roadNet-TX}, {\it roadNet-CA}, and {\it com-LiveJournal}, will have to do data replacement.
And the experimental result in Fig.~\ref{fig:dataexchange} demonstrate that with our proposed Priority data replacement policy, compared with the least recently used (LRU) replacement policy, the number of data replacement is reduced by up to $30.1\%$.

\begin{table*}[htbp]
\setlength{\tabcolsep}{12pt}
\small
\caption{Runtime (in seconds) comparison among our proposed methods, CPU, GPU and FPGA implementations.}
\label{tab:graphperf}
\centering
\begin{tabular}{lrrrrrc}
\specialrule{0.8pt}{0pt}{0pt}
 \multirow{2}{*}{Dataset} & \multirow{2}{*}{CPU} & \multirow{2}{*}{GPU \cite{XiongTCFPGA}} & \multirow{2}{*}{FPGA \cite{XiongTCFPGA}} & \multicolumn{3}{c}{Proposed Method} \\ \cline{5-7}
 & & & & w/o PIM & TCIM & Priority TCIM\\ \hline
 ego-facebook & 5.399 & 0.15 & 0.093 & 0.169 & 0.005 & 0.005 \\
 email-enron & 9.545 & 0.146 & 0.22 & 0.8 & 0.021 & 0.011 \\
 com-Amazon & 20.344 & N/A & N/A & 0.295 & 0.011 & 0.011 \\ 
 com-DBLP & 20.803 & N/A & N/A & 0.413 & 0.027 & 0.027 \\
 com-Youtube & 61.309 & N/A & N/A & 2.442 & 0.098 & 0.100 \\
 roadNet-PA & 77.320 & 0.169 & 1.291 & 0.704 & 0.043 & 0.025 \\
 roadNet-TX & 94.379 & 0.173 & 1.586 & 0.789 & 0.053 & 0.030 \\
 roadNet-CA & 146.858 & 0.18 & 2.342 & 3.561 & 0.081 & 0.047 \\
 com-LiveJournal & 820.616 & N/A & N/A & 33.034 & 2.006 & 1.940 \\ \hline
 Average & & & & & 1.0 & 1.36 \\
\specialrule{0.8pt}{0pt}{0pt}
\end{tabular}
\end{table*}

\subsection{Performance and Energy Results}

TABLE~\ref{tab:graphperf} compares the performance of our proposed in-memory TC accelerator against a CPU baseline implementation, and the existing GPU and FPGA accelerators.

One can see a dramatic reduction of the execution time in the last columns from the previous three columns. Indeed, without PIM, we achieved an average $53.7\times$ speedup against the baseline CPU implementation because of data slicing, reuse, and replacement. 
With PIM, another $25.5\times$ acceleration is obtained. 
Compared with the GPU and FPGA accelerators, the improvement is $9\times$ and $23.4\times$, respectively. It is important to mention that we achieve this with a single-core CPU and $16$ MB STT-MRAM computational array. 
With the optimized Priority data replacement policy (named as Priority TCIM), we can get another $1.36\times$ speedups.

\begin{figure}[h]
\centering
\includegraphics[width = 1.0\linewidth]{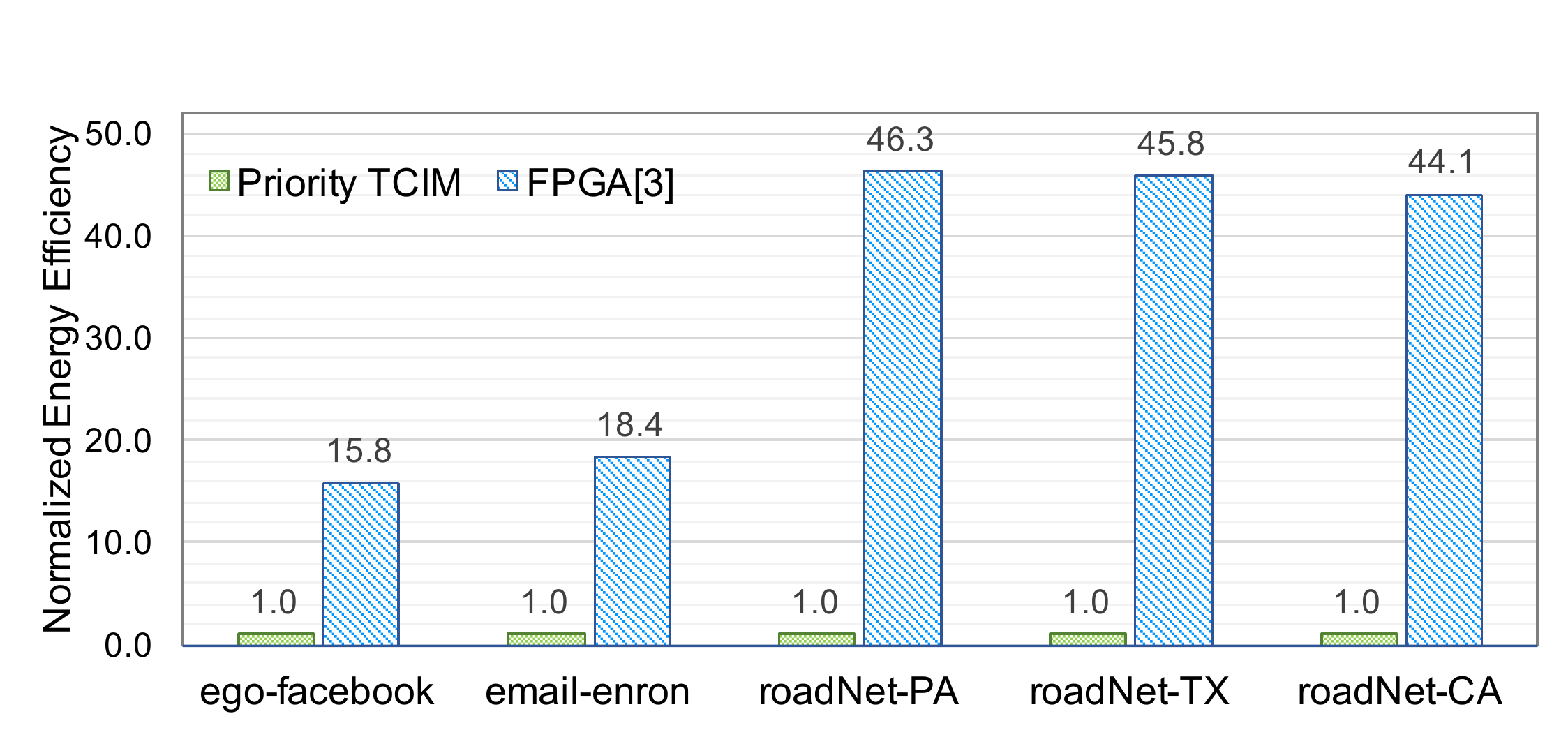}
\caption{Normalized results of energy consumption for Priority TCIM with respect to FPGA.}
\label{fig:energy}
\end{figure}

As for the energy savings, as shown in Fig.~\ref{fig:energy}, our approach has $34\times$ less energy consumption compared to the energy-efficient FPGA implementation \cite{XiongTCFPGA}, which benefits from the non-volatile property of STT-MRAM and the in-situ computation capability.

\section{Conclusion}\label{sec:conclusion}

In this paper, we propose a new triangle counting (TC) method, which uses massive bitwise logic computation, making it amenable for in-memory implementations.
We further propose a sparsity-aware processing-in-memory architecture for efficient in-memory TC accelerations. A straightforward data reuse strategy is proposed to save write operations as well as a data slicing technique to exploit sparsity in the benefit of saving even more write operations. By data slicing, the computation could be reduced by $99.99\%$, meanwhile the compressed graph data can be directly mapped onto STT-MRAM computational memory array for bitwise operations, and the proposed data reuse and replacement strategy reduces $60.5\%$ of the memory \texttt{WRITE} operations. 
Device-level simulations were carried out to obtain MTJ parameters then used in NVSim to estimate memory array performance. This, in turn, is then used by a behavioral-level simulator developed to compute energy and latency metrics.
The device-to-architecture co-simulations demonstrate that our in-memory accelerator achieves improvement in terms of speed and energy efficiency by an order of magnitude over traditional GPU/FPGA accelerators.



\section*{Acknowledgement}

Xueyan Wang's work is supported by the National Natural Science Foundation of China (No. 62004011) and State Key Laboratory of Computer Architecture (No. CARCH201917). Jianlei Yang's work is supported by National Natural Science Foundation of China (No. 62072019). Xiaotao Jia's work is supported by the Joint Funds of the National Natural Science Foundation of China (Grant No.  U20A20204).
Rong Yin's work is supported by the Special Research Assistant project of CAS (No.E0YY221-2020000702) and the National Natural Science Foundation of China (No.62106259).

\bibliographystyle{IEEEtran}

\footnotesize

\begingroup
\bibliography{TC2021}

\endgroup

\begin{IEEEbiography}[{\includegraphics[width=1in,height=1.25in,clip,keepaspectratio]{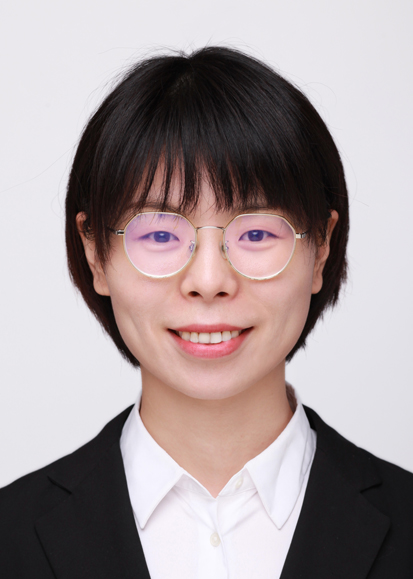}}]{Xueyan Wang}
received the B.S. degree in computer science and technology from Shandong University, Jinan, China,in 2013, and the Ph.D. degree in computer science and technology from Tsinghua University, Beijing, China, in 2018. From 2015 to 2016, she was a visiting scholar in University of Maryland, College Park, MD, USA. She is currently an Assistant Professor with the School of Integrated Circuit Science and Engineering in Beihang University, Beijing, China. Her current research interests include processing-in-memory architectures, AI chip, and hardware security.
\end{IEEEbiography}

\begin{IEEEbiography}[{\includegraphics[width=1in,height=1.25in,clip,keepaspectratio]{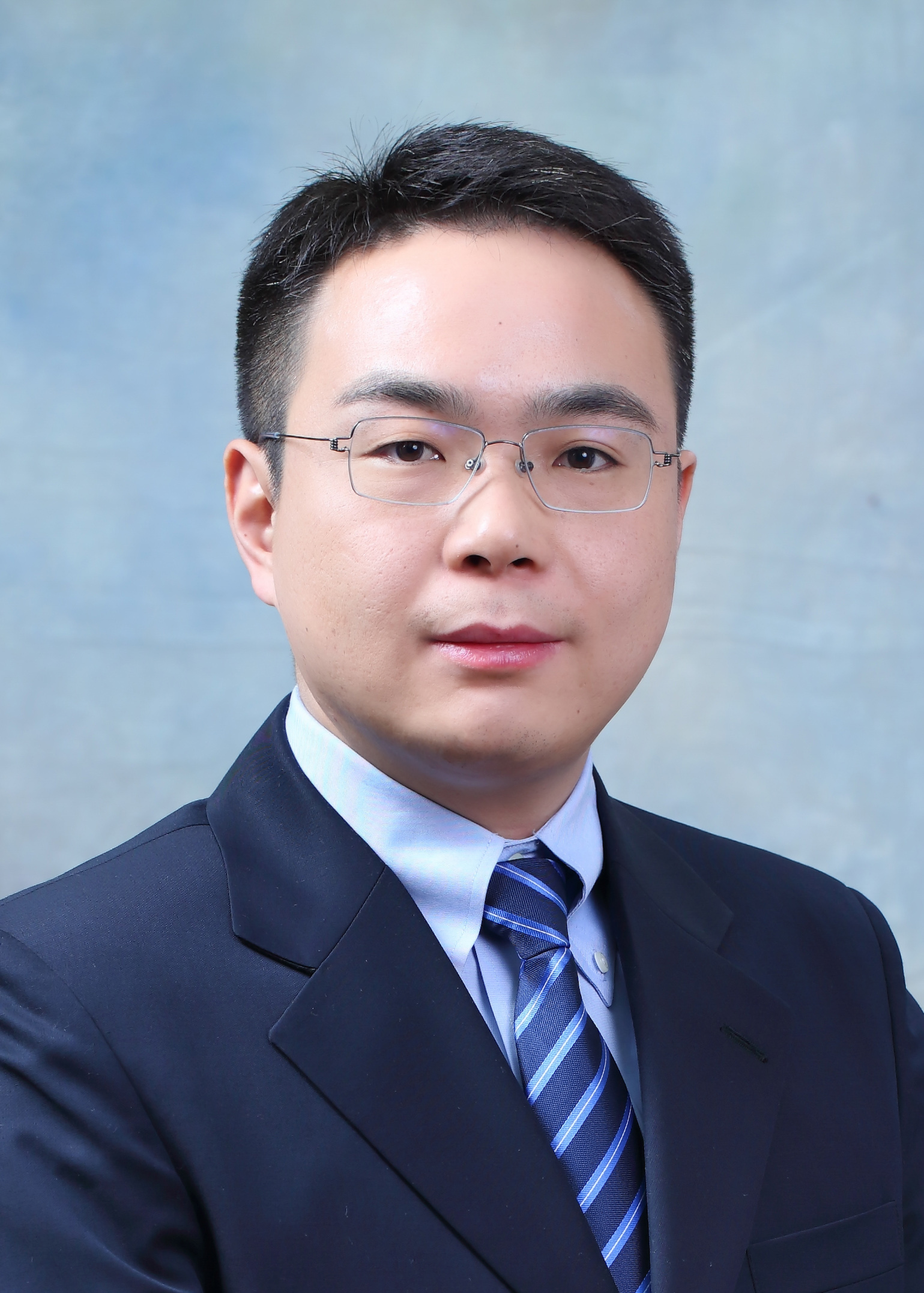}}]{Jianlei Yang}
(S'11-M'14-SM'20) received the B.S. degree in microelectronics from Xidian University, Xi'an, China, in 2009, and the Ph.D. degree in computer science and technology from Tsinghua University, Beijing, China, in 2014.
He is currently an Associate Professor in Beihang University, Beijing, China, with the School of Computer Science and Engineering. From 2014 to 2016, he was a post-doctoral researcher with the Department of ECE, University of Pittsburgh, Pennsylvania, United States.
His current research interests include computer architectures and neuromorphic computing systems.
Dr. Yang was the recipient of the First/Second place on ACM TAU Power Grid Simulation Contest in 2011/2012. He was a recipient of IEEE ICCD Best Paper Award in 2013, ACM GLSVLSI Best Paper Nomination in 2015, IEEE ICESS Best Paper Award in 2017, ACM SIGKDD Best Student Paper Award in 2020.
\end{IEEEbiography}

\begin{IEEEbiography}[{\includegraphics[width=1in,height=1.25in,clip,keepaspectratio]{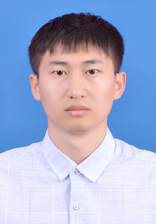}}]{Yinglin Zhao}
received the M.S. degree in software engineering from Xidian University, Xi'an, China, in 2017, and now is pursuing the Ph.D. degree in electrical engineering at School of Electronic and Information Engineering, Beihang University, Beijing, China. His research interests include the computer systems architecture and the design of non-volatile memory. 
\end{IEEEbiography}

\begin{IEEEbiography}[{\includegraphics[width=1in,height=1.25in,clip,keepaspectratio]{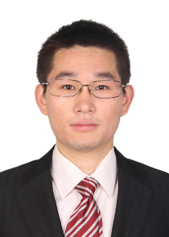}}]{Xiaotao Jia}
received the B.S. degree in mathematics from Beijing Jiao Tong University, Beijing, China, in 2011, and the Ph.D. degree in computer science and technology from Tsinghua University, Beijing, China, in 2016. He is currently an Associate Professor with the
School of Microelectronics in Beihang University, Beijing,
China. From 2016 to 2019, he was a post-doctor researcher
with the Microelectronics in Beihang University, Beijing, China.
His current research interests include spintronic circuits, stochastic computing and Bayesian deep learning.
\end{IEEEbiography}

\begin{IEEEbiography}[{\includegraphics[width=1in,height=1.25in,clip,keepaspectratio]{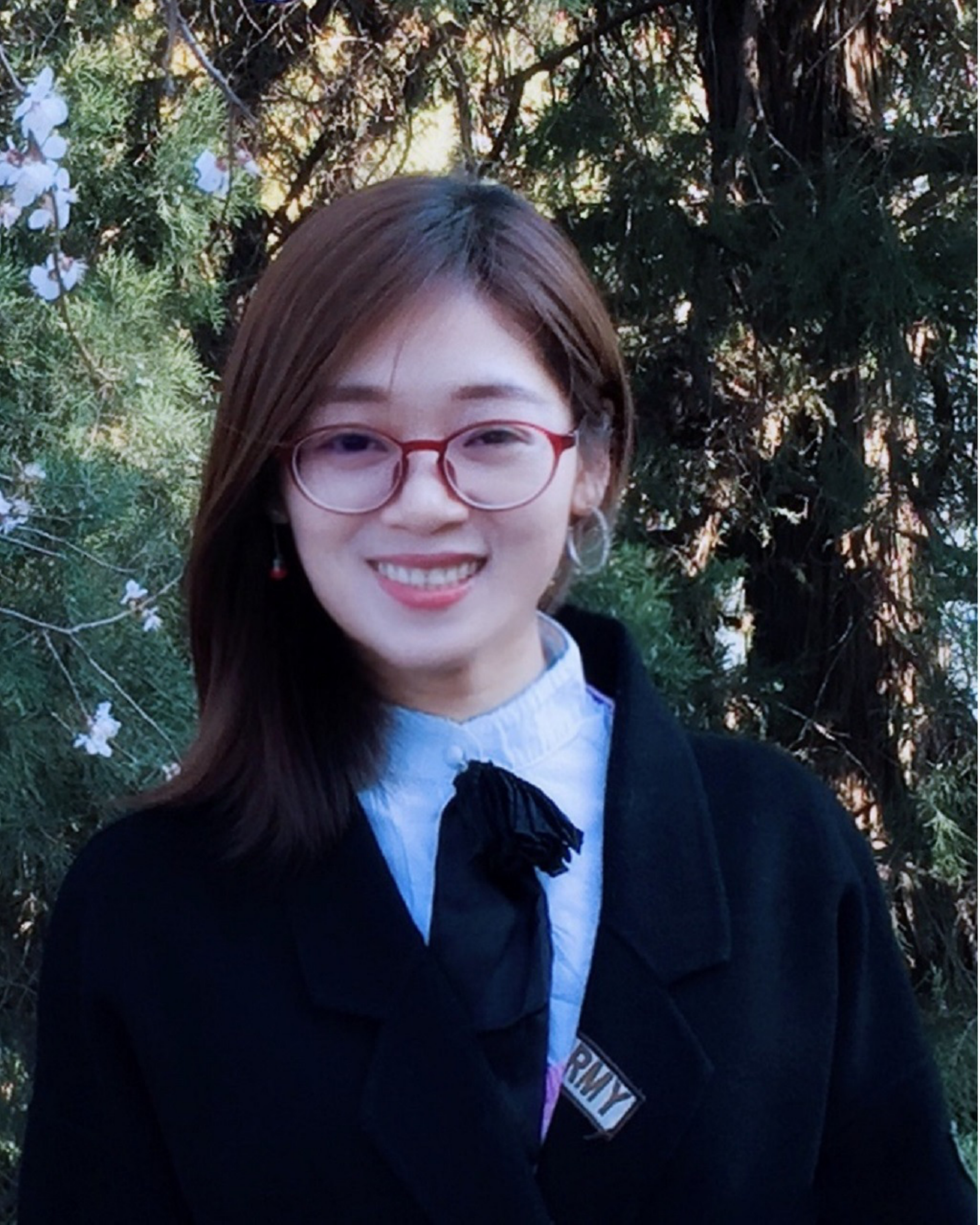}}]{Rong Yin}
received the Ph.D. degree from the Institute of Information Engineering, Chinese Academy of Sciences, Beijing, China, and the
School of Cyber Security, University of Chinese Academy of Sciences, Beijing, China, in 2020.
She is currently an Associate Professor with the Institute of Information Engineering, Chinese Academy of Sciences, Beijing, China. Her current research interests include machine learning, data mining, statistical theory, optimization algorithm, and large-scale kernel methods.
\end{IEEEbiography}

\begin{IEEEbiography}[{\includegraphics[width=1in,height=1.25in,clip,keepaspectratio]{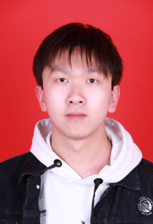}}]{Xuhang Chen}
received the B.S. degree in computer science and technology from Dalian University of Technology, Dalian, China, in 2020, and now is pursuing the M.S. degree at School of Integrated Circuit Science and Engineering, Beihang University, Beijing, China. His research interests include the graph computing accelerations with emerging in-memory computing architectures. 
\end{IEEEbiography}

\begin{IEEEbiography}[{\includegraphics[width=1in,height=1.25in,clip,keepaspectratio]{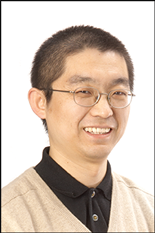}}]{Gang Qu}
(Fellow, IEEE) received the B.S. and M.S. degrees in mathematics from the University of Science and Technology of China, in 1992 and 1994, respectively, and the Ph.D. degree in computer science from the University of California, Los Angeles, in 2000. Upon graduation, he joined the University of Maryland at College Park, where he is currently a professor in the Department of
Electrical and Computer Engineering and Institute for Systems Research. Dr. Qu is the director of Maryland Embedded Systems and Hardware Security Lab and the Wireless Sensors Laboratory.
His primary research interests are in the area of embedded systems
and VLSI CAD with focus on low power system design and hardware related security and trust. He is an associate editor for the IEEE Transactions on Computer-Aided Design of Integrated Circuits and Systems, IEEE Transactions on Emerging Topics in Computing,  ACM Transactions on Design Automation of Electronic Systems, Journal of Hardware and System Security, Journal of Computer Science and Technology,  and Integration, the VLSI Journal. He has served 18 times as the general or program chair/co-chair for conferences, symposiums and workshops. He is the co-founder of  IEEE Asian Hardware Oriented Security and Trust Symposium,  Hot Picks in Hardware and System Security Workshop, and the  IEEE CEDA Hardware Security and Trust Technical Committee. 
\end{IEEEbiography}

\begin{IEEEbiography}[{\includegraphics[width=1in,height=1.25in,clip,keepaspectratio]{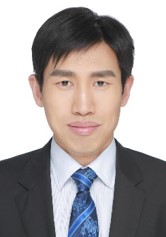}}]{Weisheng Zhao}
(Fellow, IEEE) received the Ph.D. degree in physics from
University of Paris Sud, Paris, France, in 2007. He is currently
the Professor with the School of Integrated Circuit Science and Engineering in Beihang University, Beijing, China. In 2009, he joined the French National Research Center (CNRS), as a Tenured Research Scientist. Since 2014, he has been a Distinguished Professor with Beihang University, Beijing, China. He has published
more than 200 scientifc articles in leading journals and conferences, such as Nature Electronics, Nature Communications, Advanced Materials, IEEE Transactions, ISCA and DAC. His current research interests include the hybrid integration of nano-devices with CMOS circuit and new nonvolatile memory (40-nm technology node and below) like MRAM circuit and architecture design. He is currently the Editor-In-Chief for the IEEE Transactions on Circuits and Systems I: Regular Paper.
\end{IEEEbiography}



\end{document}